\documentclass{PoS}
\usepackage{amssymb}
\usepackage{amsmath}

\title{Baseline Design for a Next Generation Wide-Field-of-View Very-High-Energy Gamma-Ray Observatory}

\ShortTitle{Design study of a wide FOV $\gamma$-ray observatory}

\author{\speaker{Harm Schoorlemmer}, Rubén López-Coto and Jim Hinton\\
        Max-Planck-Institut f\"ur Kernphysik, 69117, Heidelberg, Germany\\
        E-mail: \email{harmscho@mpi-hd.mpg.de}}

\abstract{The TeV gamma ray sky is observable by recording footprints of extensive air showers with an array of particle detectors. In the northern hemisphere there are currently two projects employing this technique: The HAWC gamma ray observatory which is currently operational in Mexico and LHAASO in the Sichuan region in China which is currently under development. In the southern hemisphere several efforts are currently ongoing to investigate the feasibility of a similar observatory at very high altitude sites in the Andes. The science case for such an observatory should be complementary to the science to be performed by the future Cherenkov Telescope Array. There are two clear directions in which such an observatory could optimize its performance. Firstly, optimize the performance of sub-TeV energies. This is especially important to provide an unbiased monitoring of a large fraction of the sky for observations of transient and extended sources. Secondly, to obtain the largest photon statistics above roughly 50 TeV, which requires a large collection area with sufficient performance in angular and energy resolution. This would enable to extend spectral measurements of Galactic sources and gives the opportunity to search for dark matter and exotic physics in a new energy range.  
Using simulated air showers and a generalized detector description the performance of a conceptual observatory is studied and the ways to optimize it will be discussed. With this approach the baseline design of such an observatory can be obtained without the need of detailed simulations of the detector hardware.}

\FullConference{35th International Cosmic Ray Conference  ICRC2017-\\
		10-20 July, 2017\\
		Bexco, Busan, Korea}

\begin{document}
%\linenumbers
\section{Introduction}
At energies above hundreds of GeV $\gamma$-rays become too rare to be detected efficiently directly by satellites. However, their interactions with the Earth's atmosphere generates a cascade of secondary particles, called Extensive Air Shower (EAS), that can be used for the determination of the properties of the primary  $\gamma$-ray. There are two different techniques to make these observations. The first of them uses optical telescopes to record the short and faint emission of Cherenkov light emitted by the charged particle within the EAS and is commonly referred to as the Imaging Atmospheric Cherenkov Telescope (IACT) technique. This method is implemented in current observatories like MAGIC \cite{MAGIC_Performance}, VERITAS \cite{VERITAS_Performance} and HESS \cite{HESS_Performance}. The second method observes the footprint of the EAS with an array of particle detectors. In order to record sufficient information about the EAS, which is needed to reconstruct properties of the primary $\gamma$-ray, these arrays are located at high altitude sites. Currently operational is the HAWC observatory \cite{HAWC_CRAB} at an altitude of 4100~m above sea-level in central Mexico. We will refer to this technique as the Air Shower Particle Detector (ASPD) technique. The work presented here is to prepare for a future observatory on the Southern Hemisphere. The science case for such an observatory, in concert with other future observatories, is discussed in \cite{SGSOScience}. \\ 
Typically IACTs perform better in accuracy on energy and direction determination compared to ASPD observatories. However, ASPD observatories have a larger instantaneous field-of-view (roughly 1\,sr) and a factor of 10 more uptime. Therefore, in some aspects, there are some clear advantages of ASPDs over IACTs:
\begin{description}
\item[Background estimation:]  The large field-of-view ensures that there is a constant sensitivity over a large fraction of the sky. Therefore an unbiased estimation of the background can be performed without the need for a special observing mode. This becomes especially important when studying emission regions that extend over several degrees on the sky.
\item[Continuous monitoring:] Roughly $\sim$2/3 of the sky is observed on a daily basis, providing long-term monitoring and possible discovery of variable sources. In addition, an ASPD-observatory can respond and issue alerts in realtime for transient events occurring within its field-of-view. 
\item[Energy reach:] Provided a sufficiently large instrumented area, the near to 100\% uptime provides for a large exposure enabling the detection of the highest energy photons. 
\end{description}

In this contribution, we study results from CORISIKA \cite{corsika} simulations of $\gamma$-ray and proton induced particles showers in the energy regime from 50~GeV to a couple of 100~TeV. Within this energy range ASPD arrays typically become efficient $\gamma$-ray detectors. In Section \ref{sec:prop} of this contribution, we briefly discuss general EAS parameters to characterize the particle footprint of the EAS and use it to motivate a baseline detector design. In Section \ref{sec:det}, we study the impact of ASPD array design choices on the anticipated performance of an observatory. We simulate an observatory as a uniform array of identical square detector units. The impact of the observatory design parameters on its $\gamma$-ray efficiency are studied. This method is used to evaluate the design of an array layout when design parameters of a single unit are chosen. Design choices can be evaluated, and tuned, the need for a full hardware simulation. The goal is that these studies are used in making design decisions for a next generation ASPD observatory in the Southern Hemisphere.

%------------------------------------------------------------------------------------------------
\section{Properties of $\gamma$ and proton induced air showers}
\label{sec:prop}
%%text copied from draft paper
The vast majority of the energy within the air showers is carried to the ground by electrons and positrons (hereafter electrons), photons, positive and negative muons (hereafter muons). The combined energy carried by the electrons and photons we call the electromagnetic energy $E_{\text{em}}$ throughout this work. Although most ASPDs measure charged particles, there is usually a conversion layer to transform energy from high energy photons, through pair-production, into electrons. The photons and electrons typically loose a significant, if not all, their energy in the detection medium  and therefore the observed signal is proportional to $E_{\text{em}}$. The muons, in contrast, typically do not lose a significant amount of their energy in the detection medium and are therefore counted by number. The efficiency and purity with which muons can be counted depends a lot on detector design choices. 

\begin{figure}[htbp]
   	\centering
   	\includegraphics[width=0.7\textwidth]{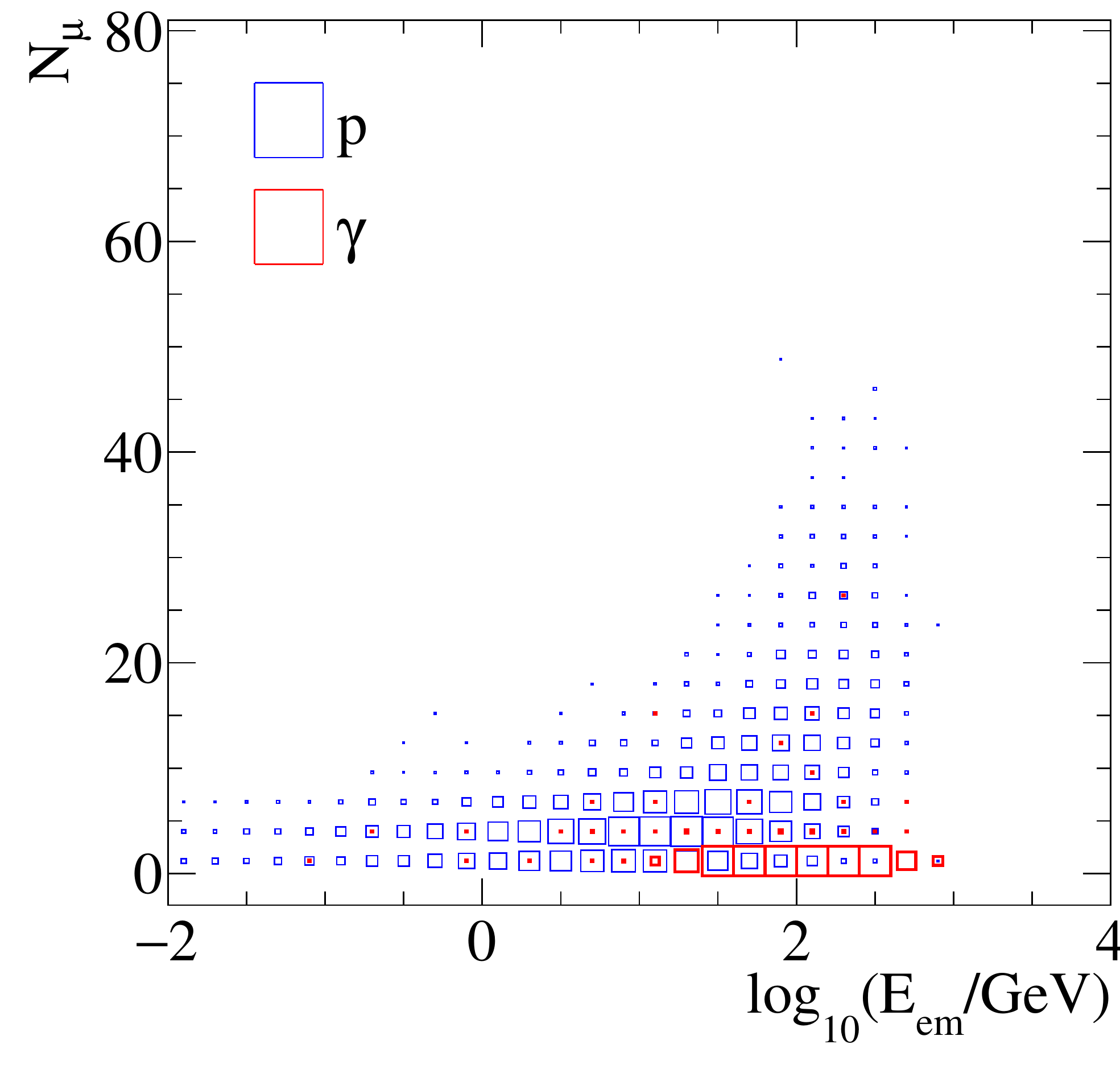} 
%	 \includegraphics[width=0.56\textwidth]{EmDist_50.pdf} 		
%	\caption{Distributions of the number of muons as a function of energy in the electromagnetic shower as observed at ground level within 100\,m from the impact point. The showers to generate these distributions had a primary energy of 1 TeV and a vertical incoming direction. The ground level is at 5000\,m a.s.l.. The sizes of the squares scale logarithmic with the density of the distributions.}
	\caption{Distributions of the number of muons as a function of energy in the electromagnetic shower as observed at ground level within 100\,m from the impact point. The showers to generate these distributions had a primary energy of 1 TeV and a vertical incoming direction. The ground level is at 5000\,m a.s.l.. The sizes of the squares scale linearly with the density of the distributions.}
   	\label{fig:dists}
\end{figure}
Figure \ref{fig:dists} shows the distribution of proton and $\gamma$-ray induced EASs in the number of muons per shower $N_\mu$ and $E_{\text{em}}$. They arrive from zenith and have an energy of 1~TeV. Clearly illustrated is that proton and $\gamma$-ray induced air showers can be separated using the number of muons within the air shower. This feature will be used in the second part of the contribution, where ASPD observatories will be equipped with muon detectors and the number of observed muons will be used to reject proton induced air showers.
 Also clear from Figure \ref{fig:dists} is that for $\gamma$-ray and proton induced air showers with the same initial energy, the amount of electromagnetic energy that reaches the ground, $E_{\text{em}}$, is on average significantly lower for proton induced EASs. In this case, the majority of the energy went into the production of muons. 

%\begin{figure}[htbp]
%   \centering
%   	\includegraphics[width=0.49\textwidth]{par_proton.pdf} 
%        \includegraphics[width=0.49\textwidth]{par_gamma.pdf}	
%   \caption{Dependence on the distributions of fraction of the energy of the particle that reaches ground level as electric magnetic energy as a function of slant depth. Shown are  $\mu$ and $\sigma$. Protons are show in the left panel (in blue) and the $\gamma$-ray are shown in the right panel (in red). (Maybe not show??)}
%   \label{fig:example}
%\end{figure}
%------------------------------------------------------------------------------------------------

%------------------------------------------------------------------------------------------------
\section{Uniform detector arrays:  performance and trade-offs}
\label{sec:det}
\subsection{Observatory layout}
The performance of a $\gamma$-ray observatory depends on the total amount of $\gamma$-rays recorded, the efficiency with which the background of EASs induced by different particle types can be rejected, and the resolution on direction and energy. 
In this section we study the impact of the following design parameters on the performance of an observatory:
\begin{description}
\item[Array size:] The square area in which units are placed. Default value {\bf{200\,m}$\times$200\,m}
\item[Array fill factor:] The fraction of the {\it{array size}} that is covered with detector units. Default value {\bf{75\%}}.
\item[Elevation:] The elevation of the site on which the array is built. Default value {\bf{5000\,m}}
\item[Unit size:] The sensitive area of a detector unit. Default value {\bf{4\,m$\times$4\,m}}
\item[Unit threshold:] The threshold in electromagnetic energy at which a unit triggers. Default value {\bf{20\,MeV}}
\item[Trigger Multiplicity:] The number of units that have a value above the {\it{unit threshold}}. Default value {\bf{20 units}}.
\end{description}
The default values above give the parameters of a reference observatory which we keep constant while varying one of the other parameters. 

\subsection{Performance parameters}
To study the performance of an observatory we distribute the location of the simulated EASs uniformly over the array with a direction coming from a 20 degree angle with respect to zenith. For these EASs we reconstruct the incoming direction using a likelihood fit. The input to the direction fit is the time of the first particle that hit a detector unit that passes the {\it{unit threshold}}. The hit times are compared to a model that is obtained from carefully stacking the hit times as a function of distance to the shower axis, $r$, for a large set of simulations. The timing model is binned in 20 logarithmic steps of the parameter $E_{50}$ which is the average energy per detector unit at  50~m from the shower core. %In the left panel of Figure \ref{fig:example}, we show an example of the likelihood direction fit.
 From the direction reconstruction we determine the angular resolution $\sigma_{68\%}$ as the 68\% quantile of the angular difference between true and reconstructed incoming directions. This direction reconstruction method has not been fine tuned, therefore a different method might perform better. However, we tried more simplistic direction reconstruction methods, where the achieved accuracy was poorer but similar trends were observed.\\
The $\gamma$-ray detection efficiency $\epsilon_{\gamma}$ is estimated from the fraction of events that passed the trigger conditions. The trigger condition is having enough detectors that were above their {\it{unit threshold}} to meet the {\it{trigger multiplicity}}.

In addition, we calculate the proton efficiency $\epsilon_p$ using the same conditions as in the $\gamma$-ray efficiency and an additional cut on the number of detected muons $N_{\mu}$ (each unit is assumed as being a perfect muon counter, and no cut on the muon energy is applied). 
\begin{figure}[htbp]
   \centering
   	\includegraphics[width=0.45\textwidth]{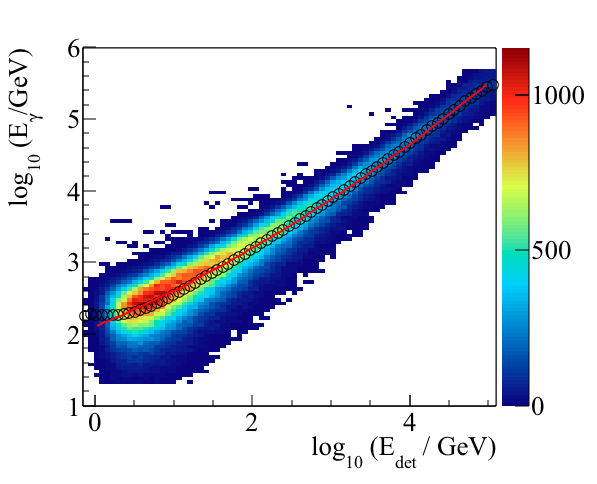} 
        \includegraphics[width=0.45\textwidth]{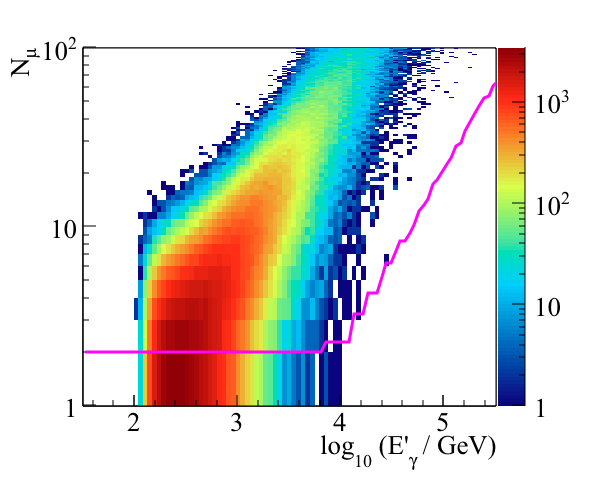}	
   \caption{Left Panel: Relation of total energy deposited in the array $E_{\text{det}}$ and the energy of the primary $\gamma$-ray $E_{\gamma}$. The open circles represent the mean values of $E_{\gamma}$, while the full distribution is shown in the color coding. The redline shows the result of linear fit to the relation. To reduce the influence of partial contained events, the distributions is obtained for showers that have a core location within $d/2$ from the center of the array, where $d$ is the length of the side of a square $d\times d$ array.  Right panel:  The distribution of the number of muons per equivalent $\gamma$-ray energy $E'_{\gamma}$ for proton showers is shown in color coding. The magenta line shows the cut value above which proton showers are rejected (See text for more detail).}
   \label{fig:example}
\end{figure}
For each proton induced EAS we calculate an equivalent energy of a $\gamma$-ray shower $E'_\gamma$ using a fitted relationship between the detected electromagnetic energy $E_{\text{det}}$ and energy of the primary $\gamma$-ray $E_\gamma$ as shown in the left panel of Figure \ref{fig:example}. For $\gamma$-ray simulations,  we calculate the value of $N_{\mu,\text{cut}}$ below which 90\% of the $\gamma$-ray simulations have a smaller value of $N_{\mu}$. If  $N_{\mu,\text{cut}} < 2$  then we set it to two, this to reflect that most likely at least two muons need to be associated with an air shower before we can use it to discriminate.  We reject showers as proton like when they are above that value of 90\% $\gamma$-ray efficiency, an example of this cut is shown in the right panel of Figure \ref{fig:example}. The proton efficiency is defined by the fraction of showers that have detected number of muons below the $N_{\mu,\text{cut}}$ and fulfill the same cuts as for the $\gamma$-ray efficiency.

%\begin{figure}[htbp]
%   \centering
%   	\includegraphics[width=0.49\textwidth]{exampleFit.png} 
%        \includegraphics[width=0.43\textwidth]{EnergyRelation.png}	
%   \caption{Left Panel: Example of the result from the likelihood direction reconstruction. The contours give the log-likelihood surface for the direction that maximizes the likelihood for the locations of the detector hits (represented by the black data points).  Right: Relation of total energy deposited in the array $E_{\text{det}}$ and the energy of the primary $\gamma$-ray $E_{\gamma}$. The open circles represent the mean values of $E_{\gamma}$, while the full distribution is shown in the color coding. The redline shows the result of linear fit to the relation. To reduce the influence of partial contained events, the distributions is obtained for showers that have a core location within $d/2$ from the center of the array, where $d$ is the length of the side of a square $d\times d$ array.   }
%   \label{fig:example}
%\end{figure}

 \subsection{Results}	
\begin{figure}[!ht]
   \centering
   	\includegraphics[width=0.32\textwidth]{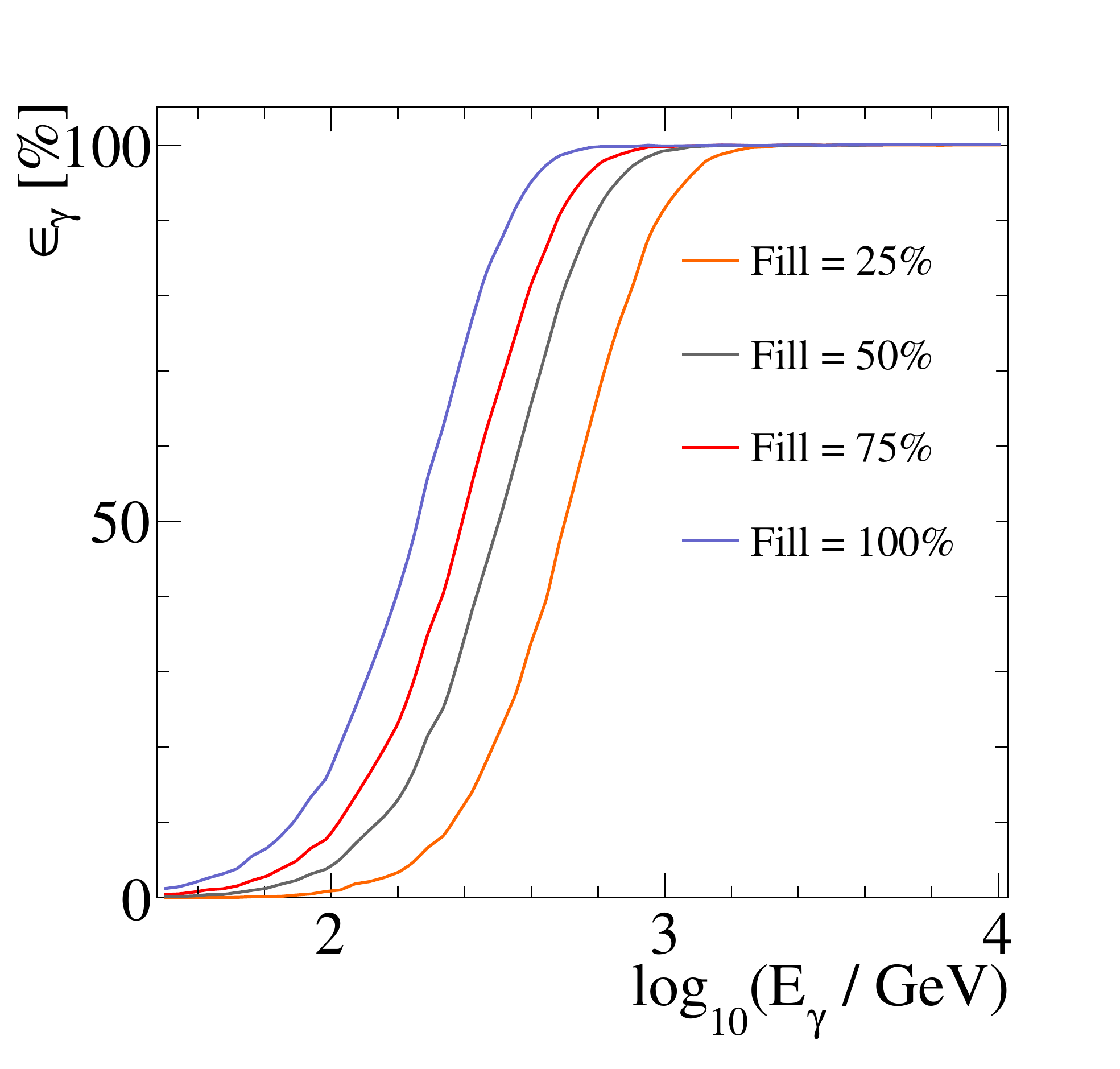} 
        \includegraphics[width=0.32\textwidth]{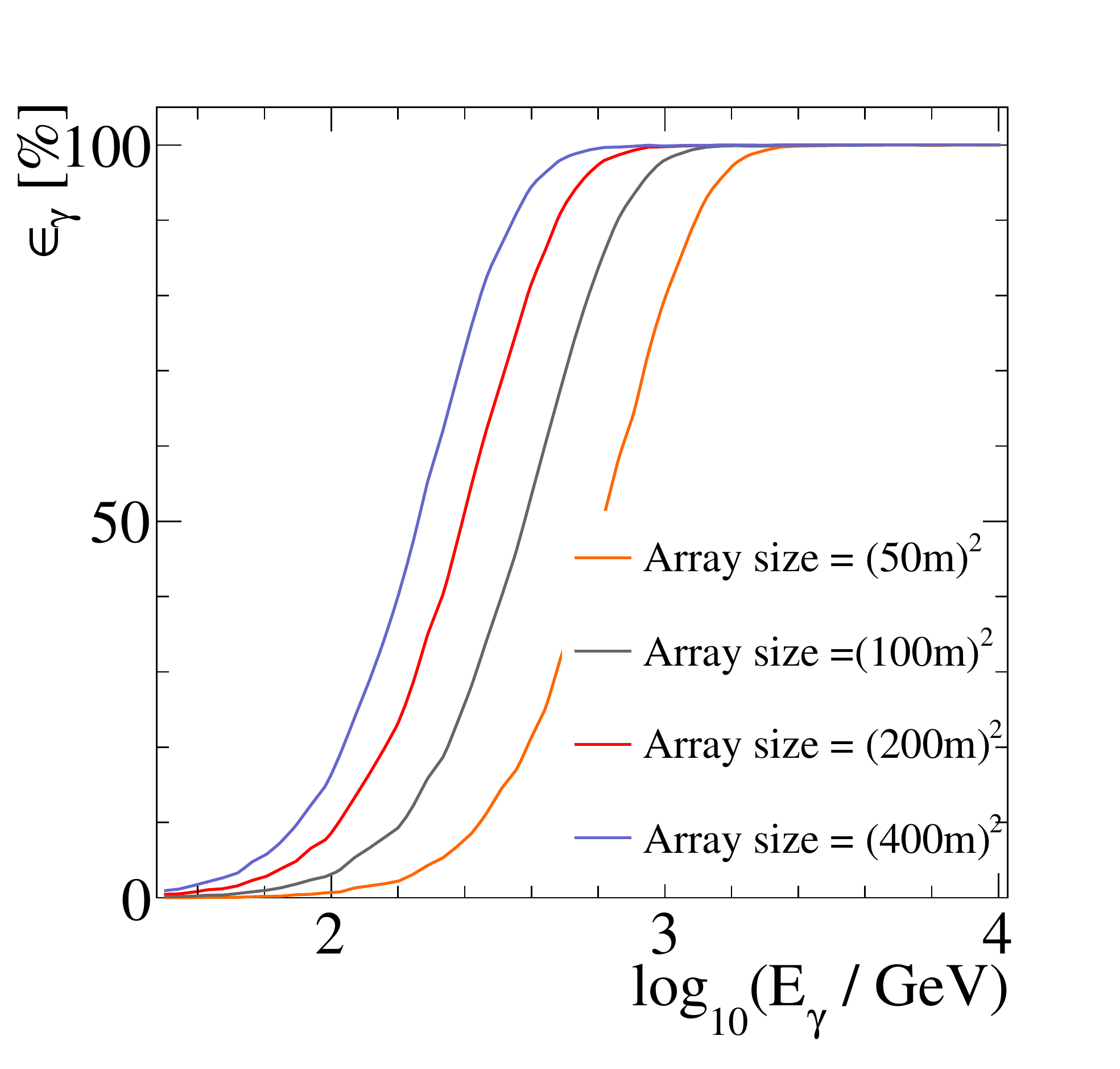}
        \includegraphics[width=0.32\textwidth]{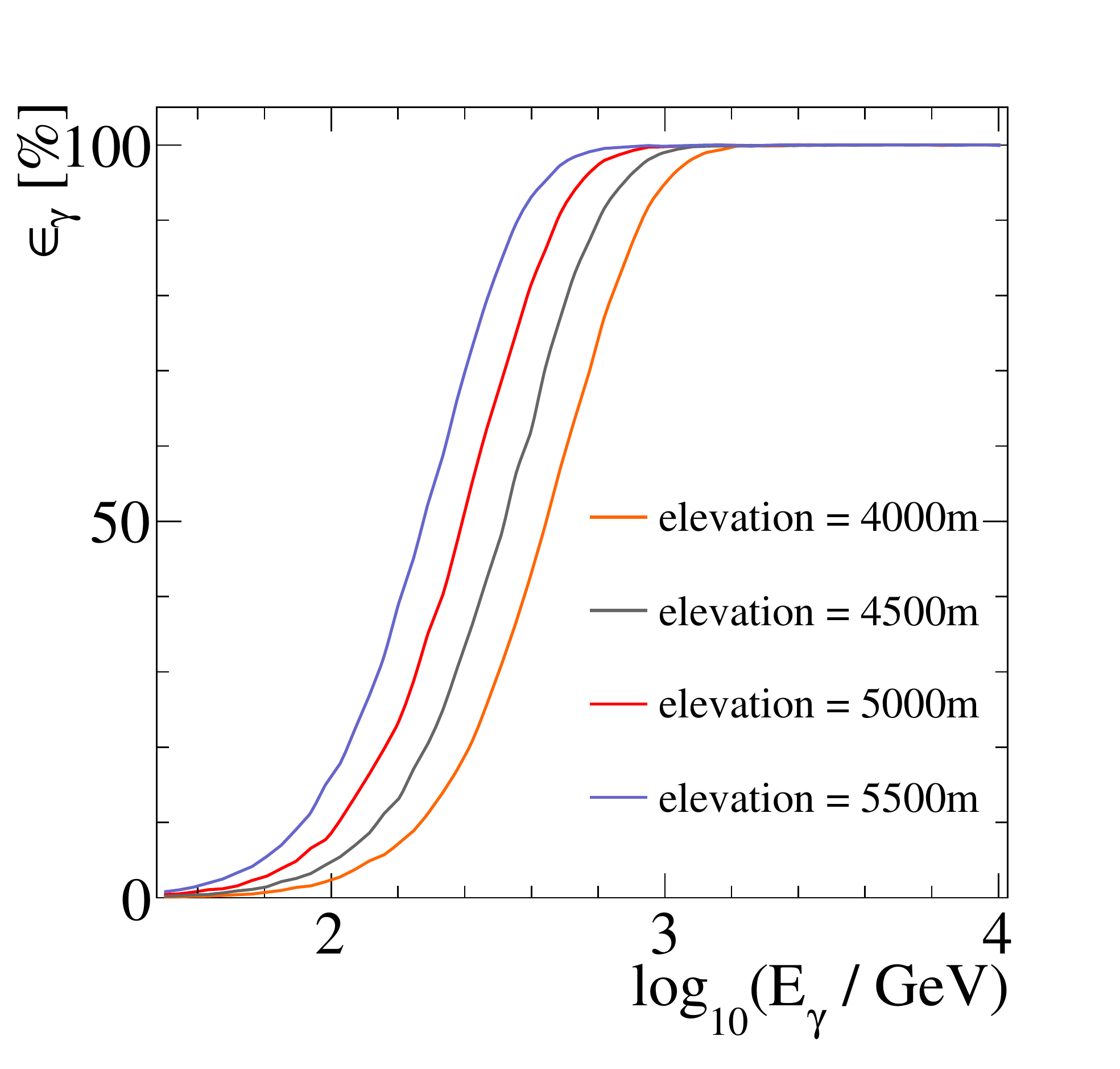}
        \includegraphics[width=0.32\textwidth]{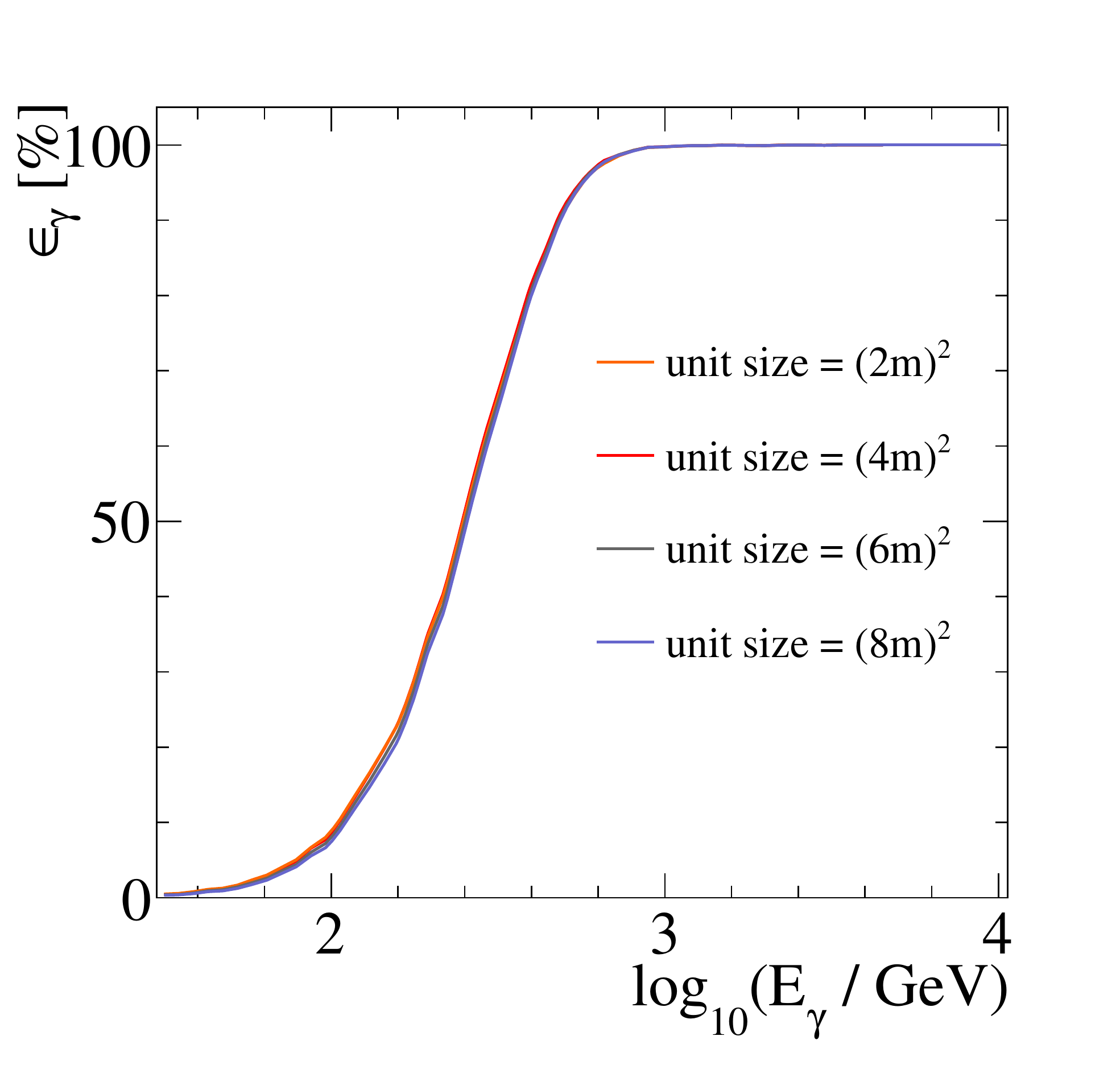}      
        \includegraphics[width=0.32\textwidth]{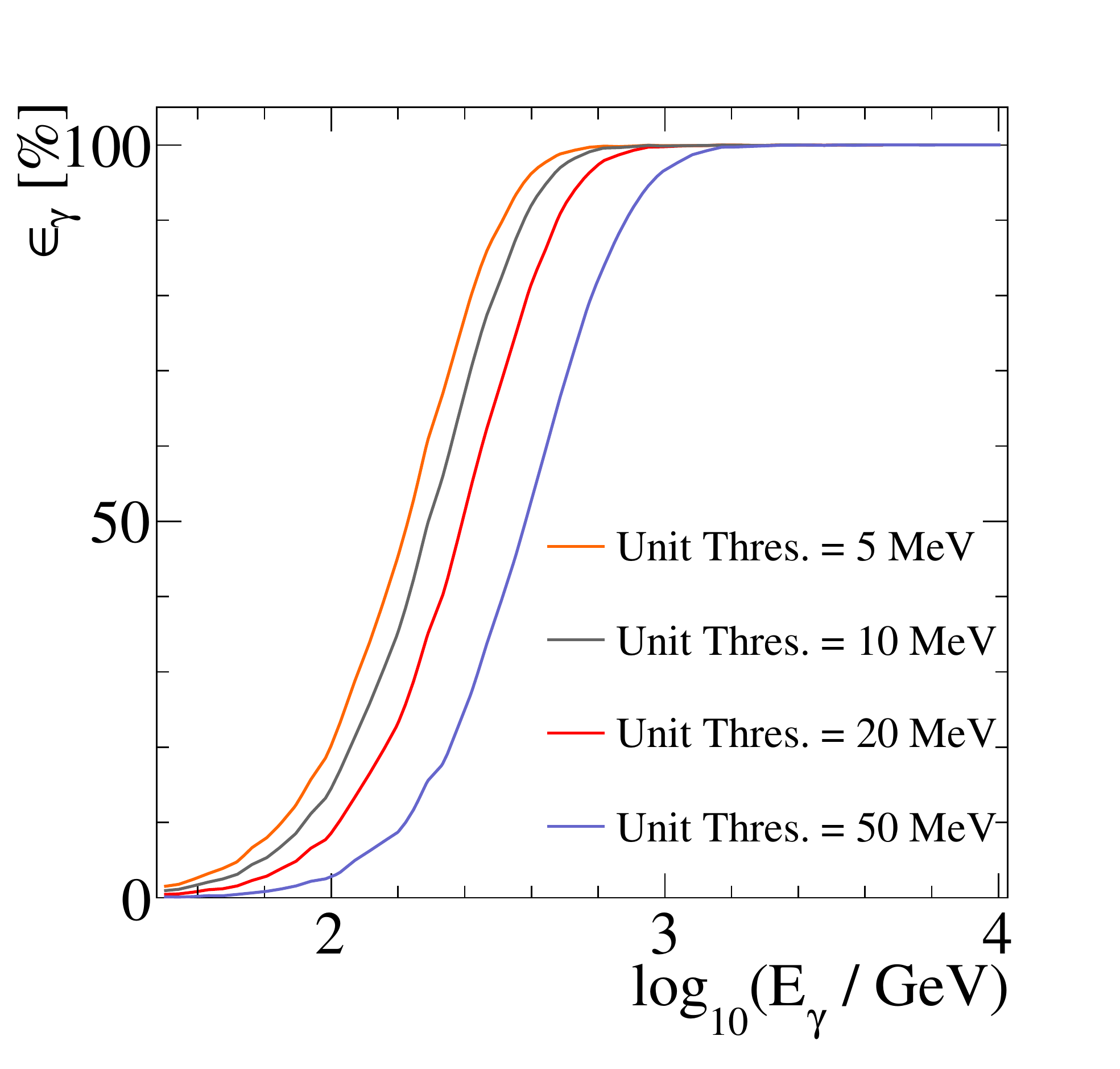}
        \includegraphics[width=0.32\textwidth]{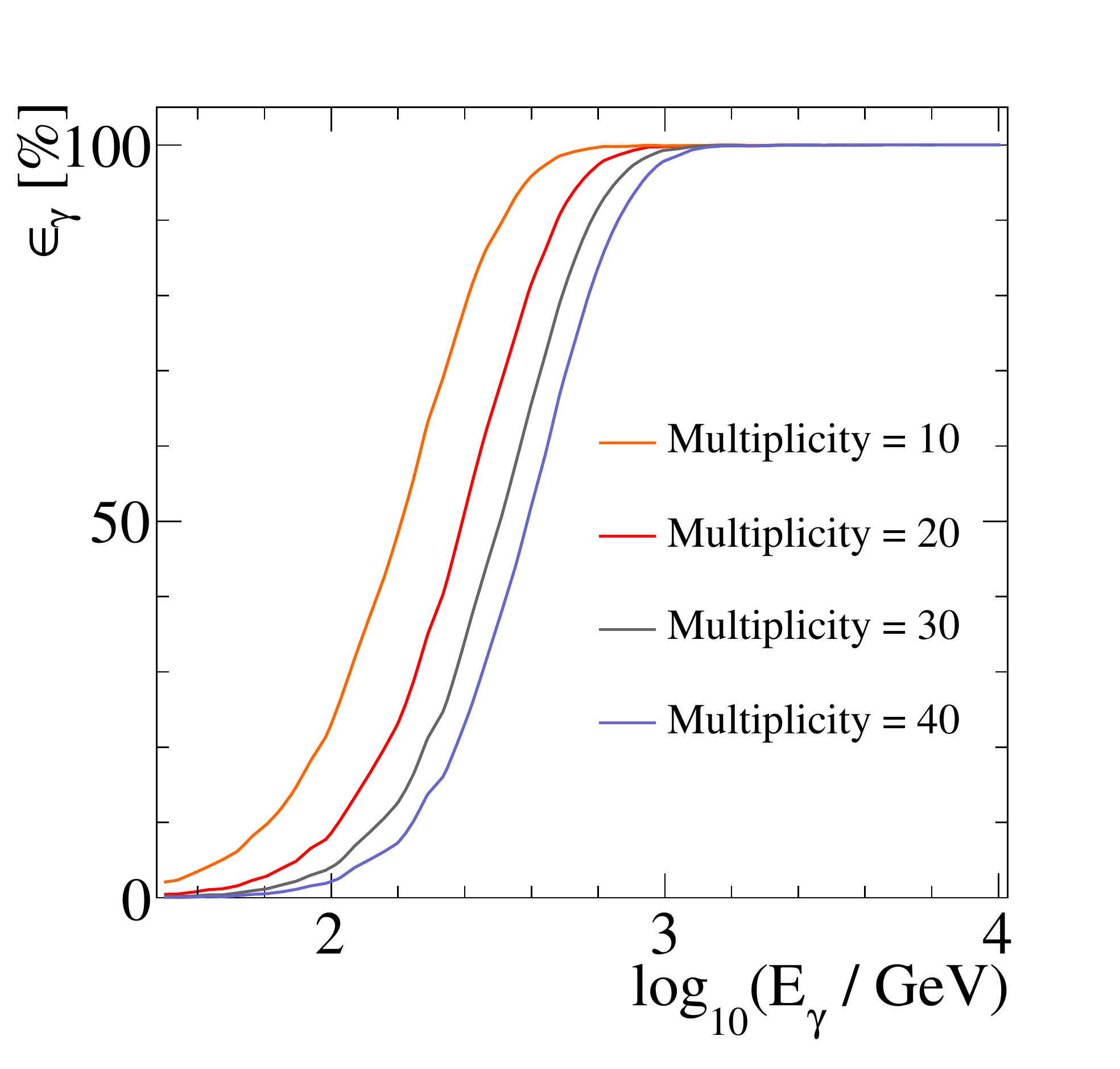}
   \caption{Efficiency for $\gamma$-ray detection with different design choices}
   \label{fig:gameff}
\end{figure}

\begin{figure}[!ht]
   \centering
	\includegraphics[width=0.32\textwidth]{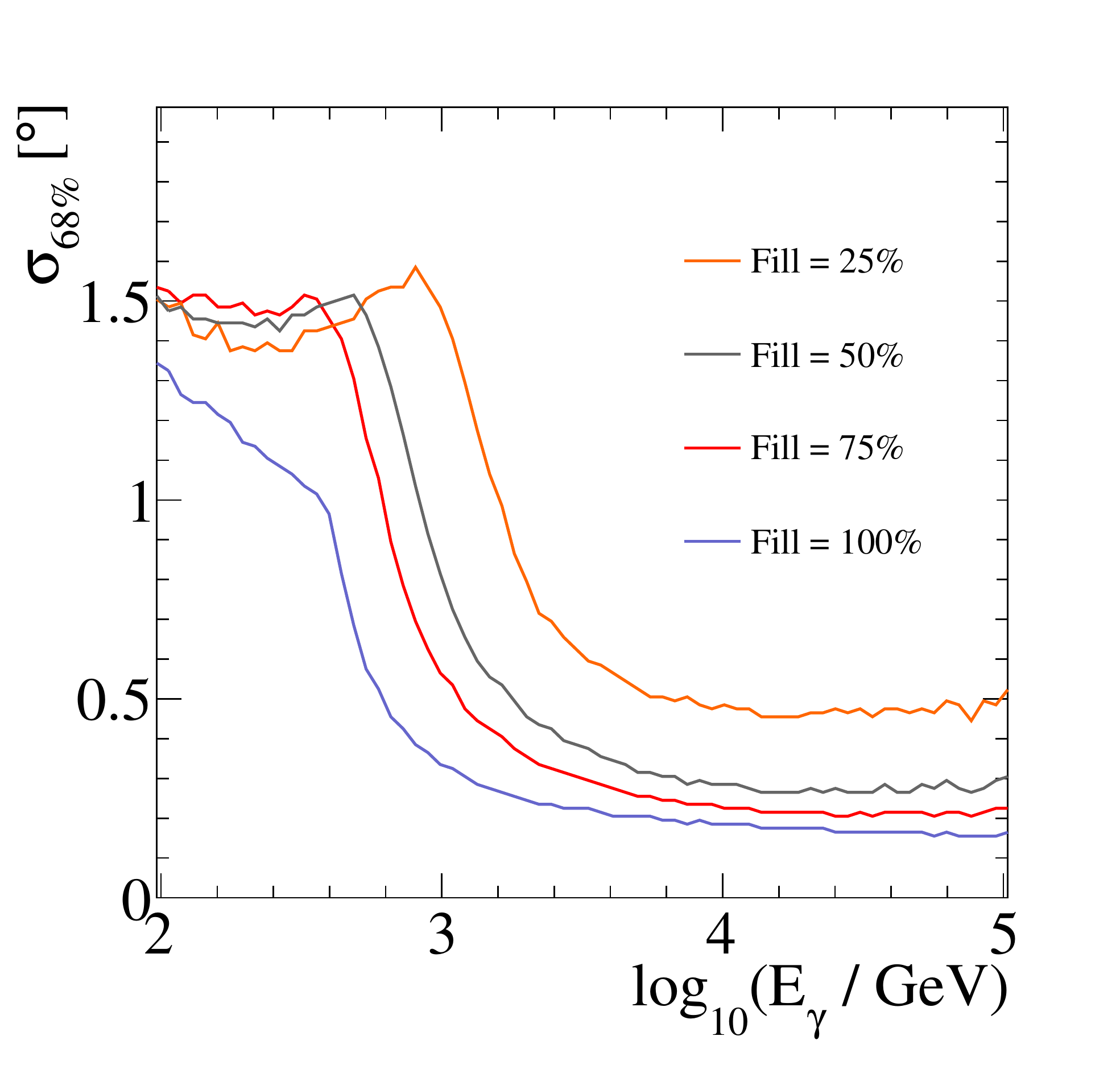}		
        \includegraphics[width=0.32\textwidth]{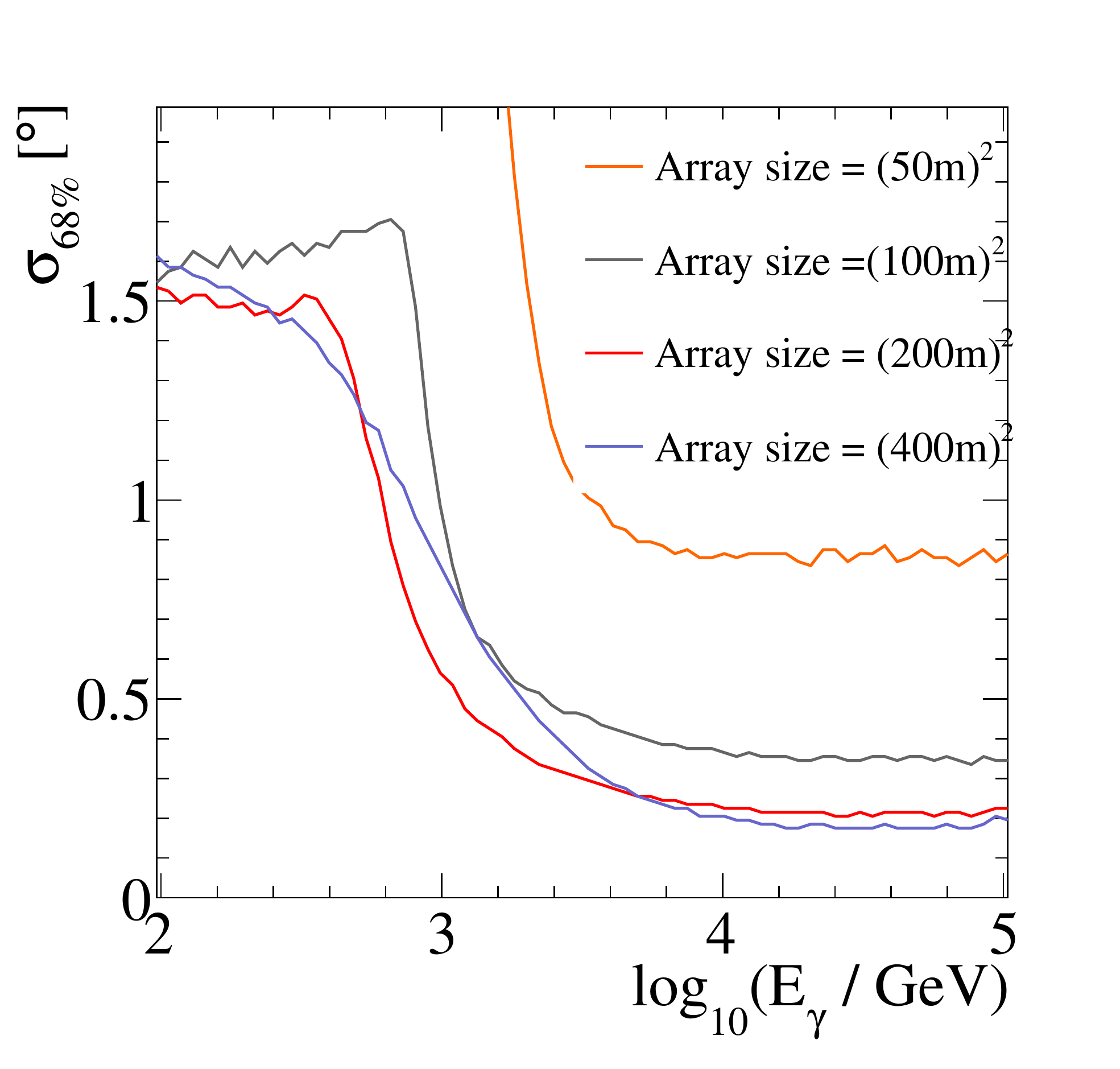}	
        \includegraphics[width=0.32\textwidth]{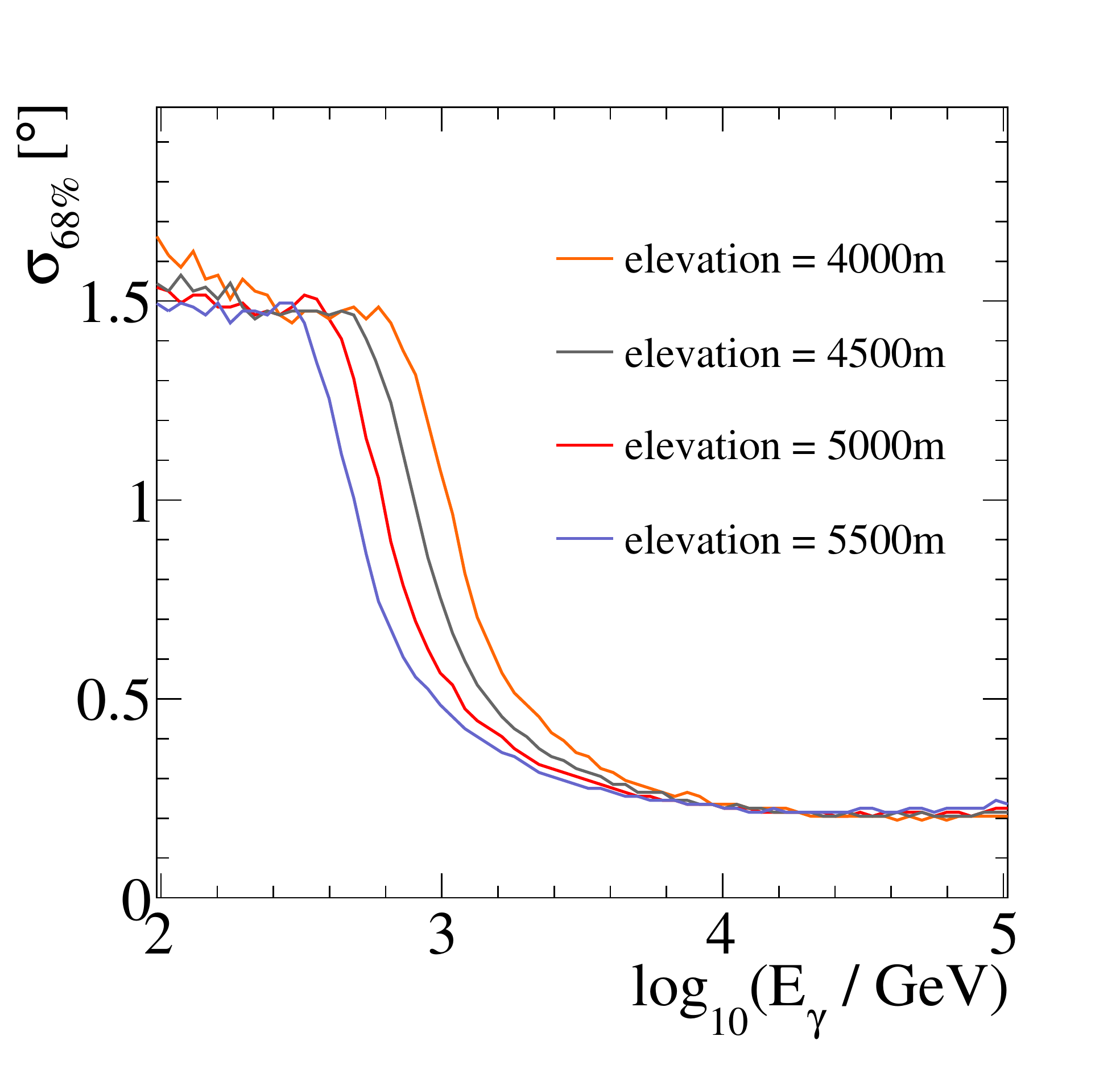}	
        \includegraphics[width=0.32\textwidth]{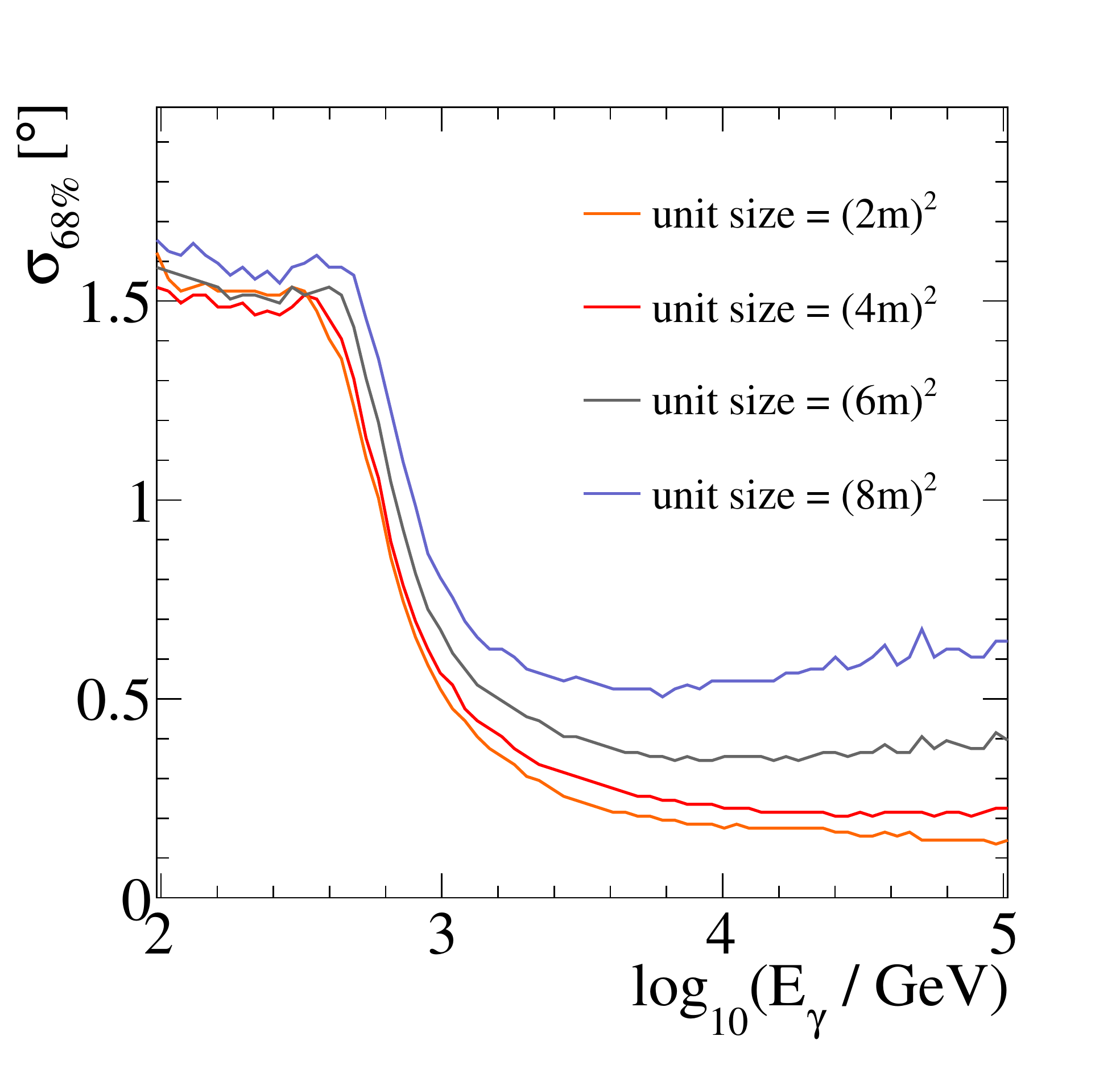}	
        \includegraphics[width=0.32\textwidth]{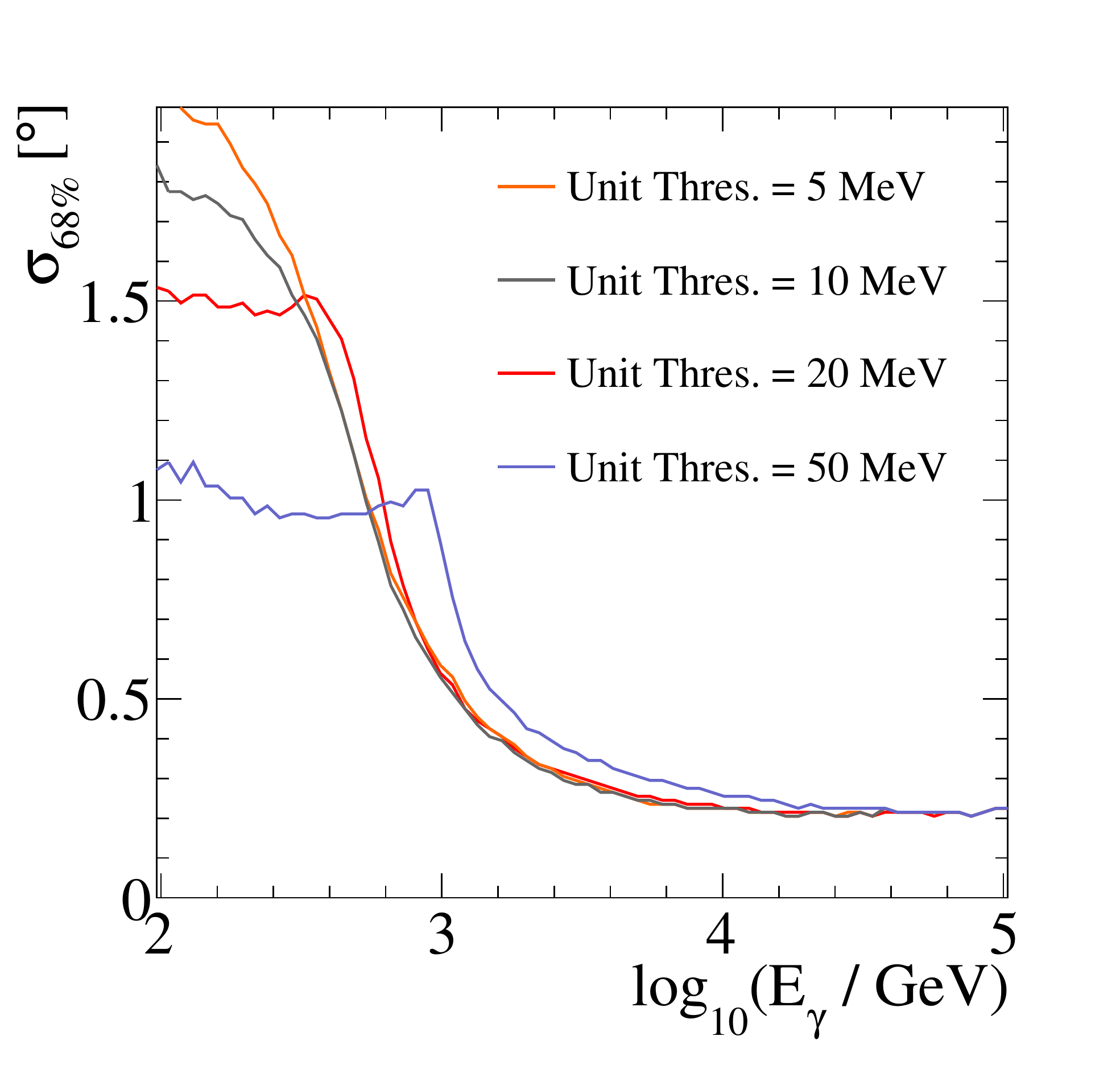}	
        \includegraphics[width=0.32\textwidth]{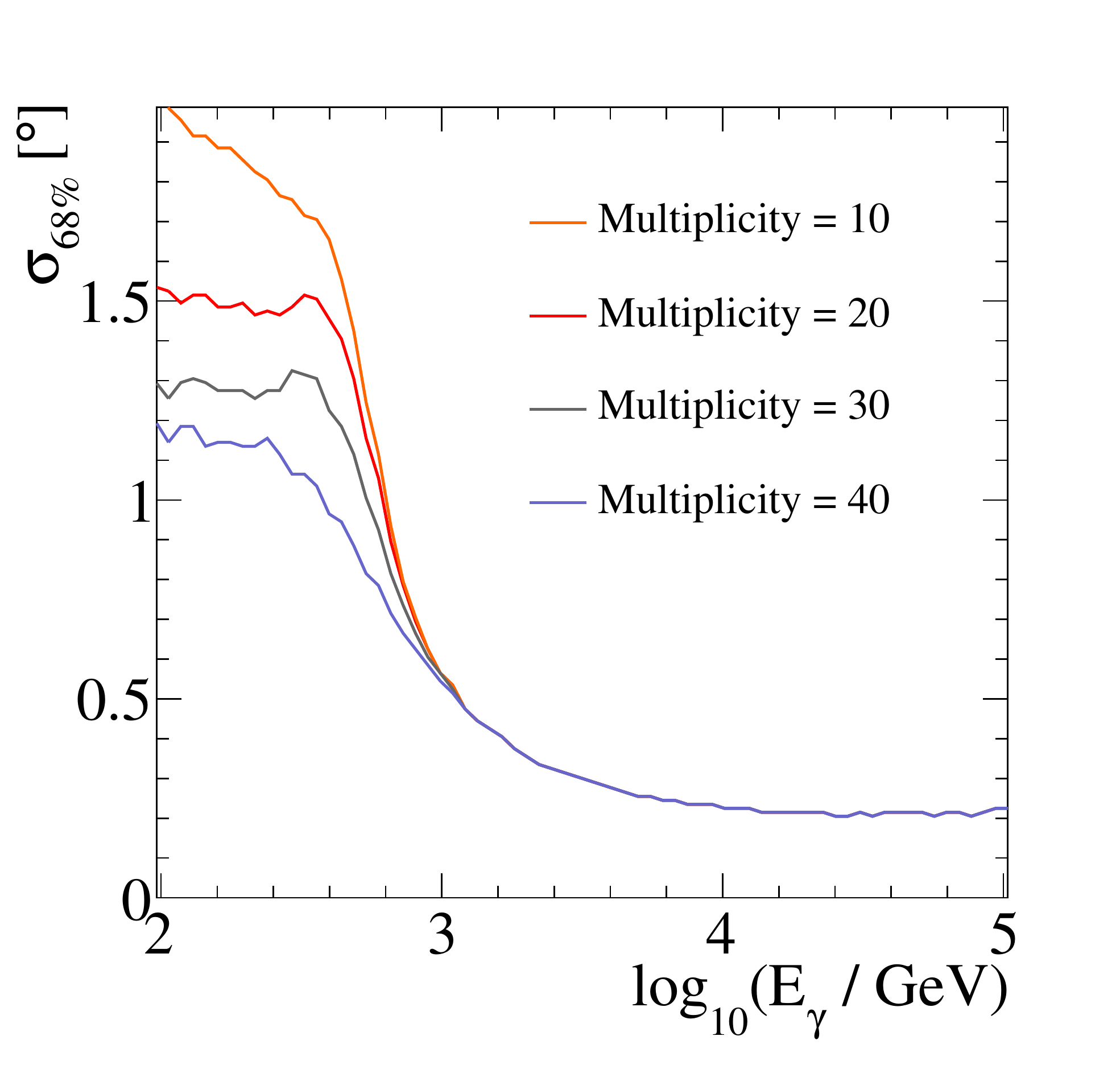}	
   \caption{Angular resolution for $\gamma$-ray detection with different design choices}
   \label{fig:angres}
\end{figure}

\begin{figure}[!ht]
   \centering
        \includegraphics[width=0.32\textwidth]{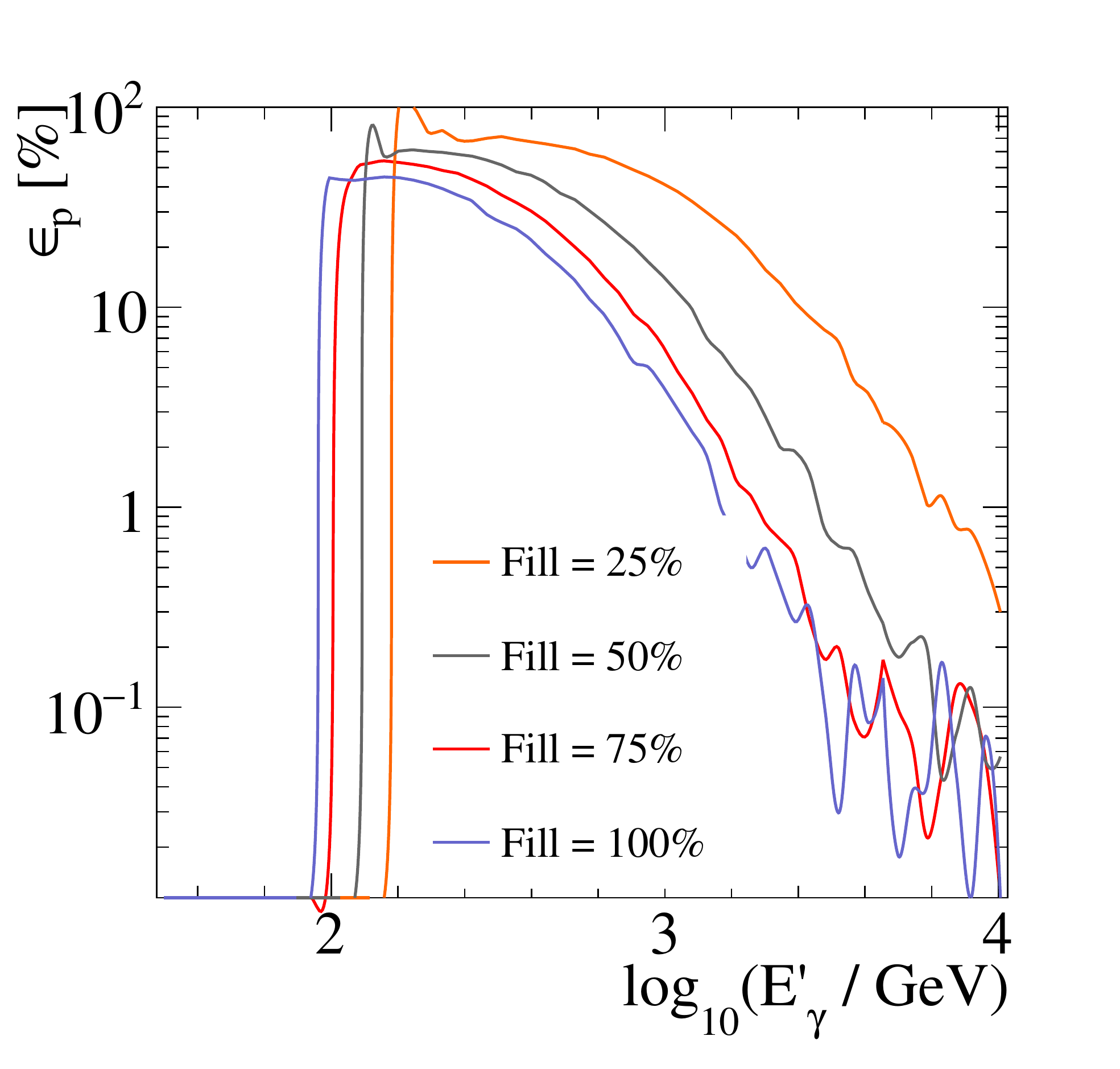}	
        \includegraphics[width=0.32\textwidth]{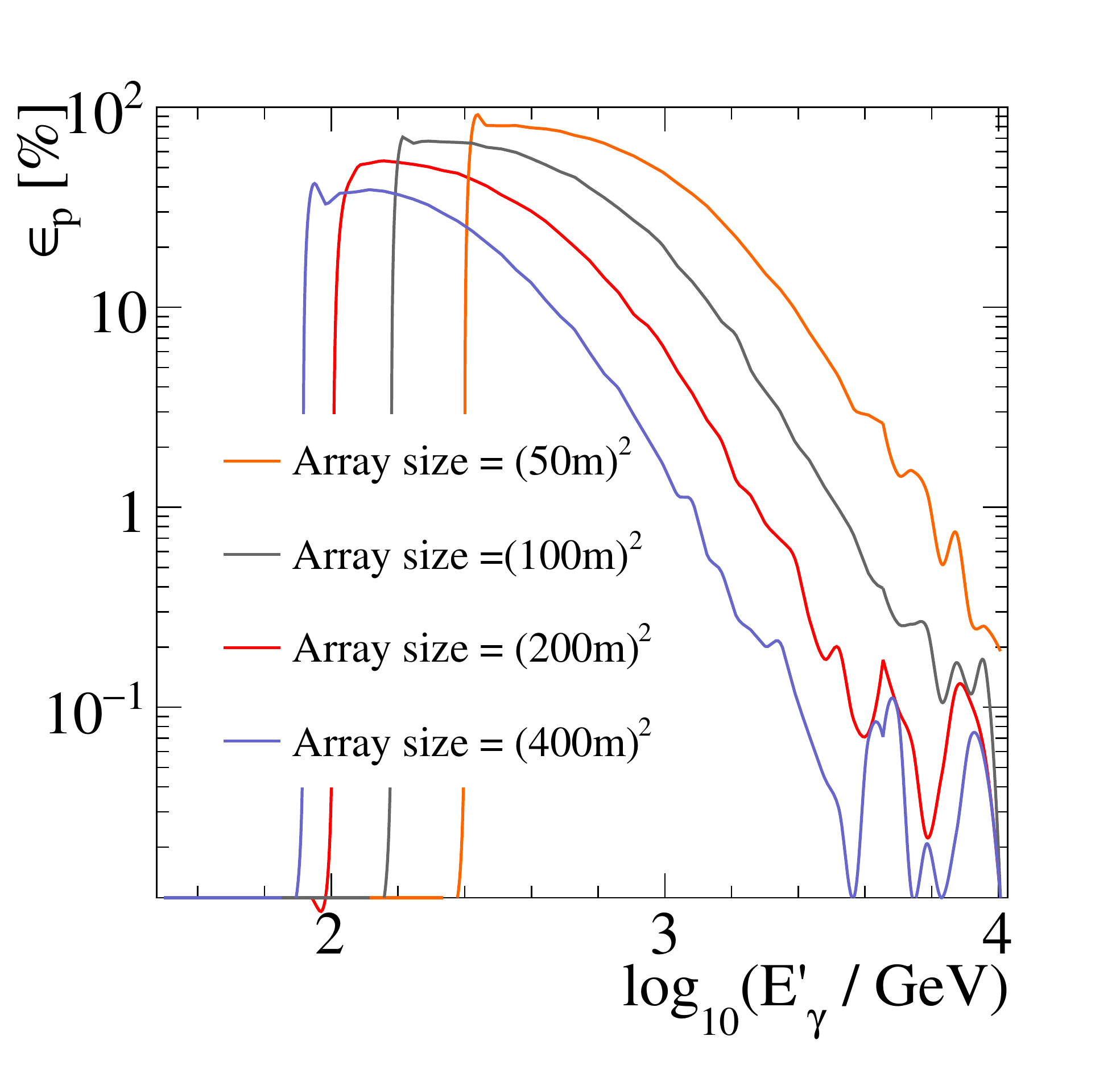}	
        \includegraphics[width=0.32\textwidth]{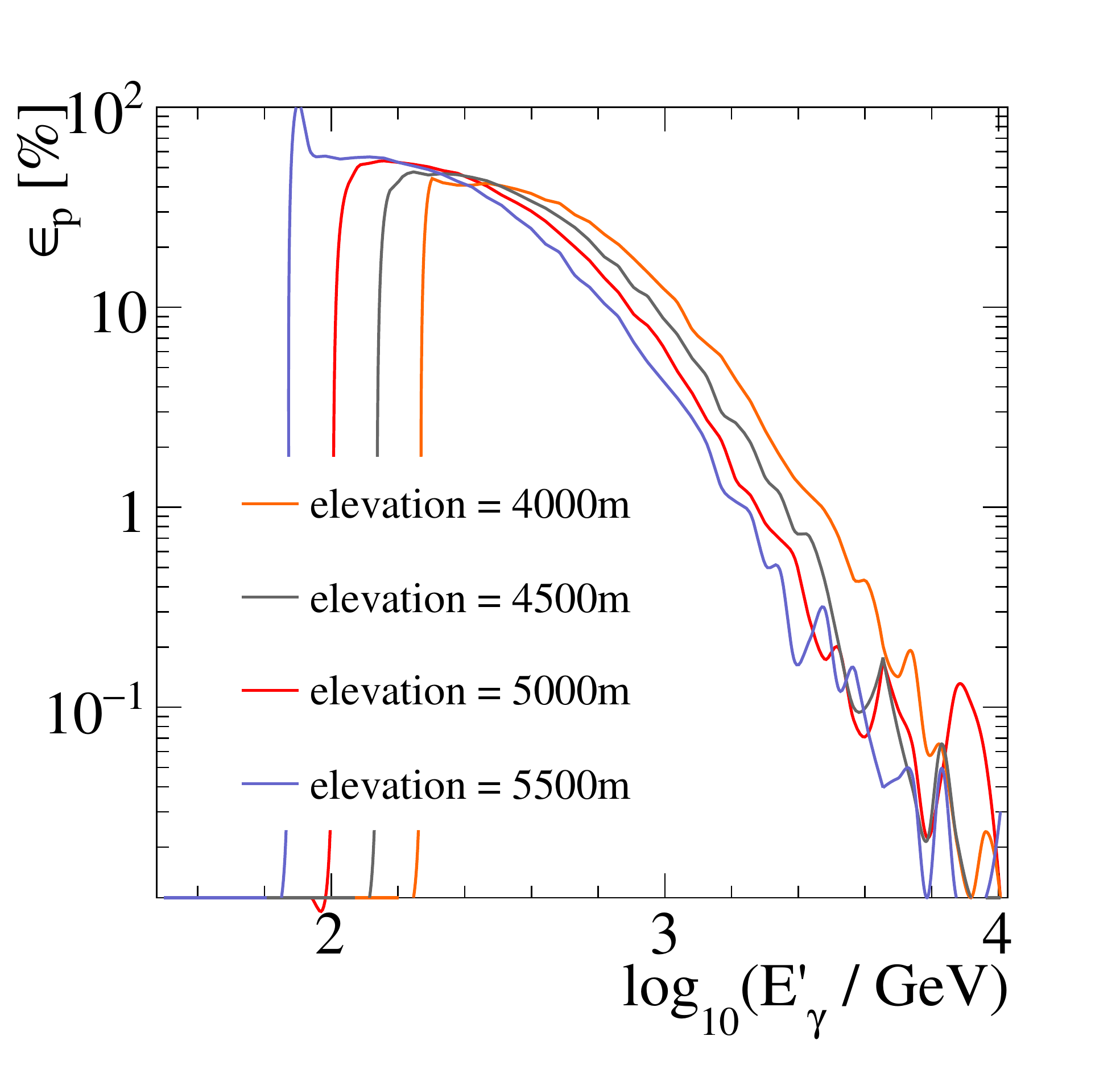}	
      \caption{Efficiency to detect protons after a cut on the number of detected muons.}
   \label{fig:proeff}
\end{figure}

Figures \ref{fig:gameff}, \ref{fig:angres}, and \ref{fig:proeff} show the results of varying the observatory design properties for $\gamma$ -ray efficiency, angular resolution, and proton efficiency respectively. 
 In each figure, the red line shows the same reference observatory.  These figures can be used to study the trade-off on performance for different observatory design. 
 The sudden cut-off at lower $E'_\gamma$ values in Figure \ref{fig:proeff} is an artifact of the way the equivalent $\gamma$-ray energy is assigned. The values just above that should be interpreted as the average value for the lower $\gamma$-ray energies. The only panels shown in Figure \ref{fig:proeff} are for the design parameters that had a significant influence on the number of detected muons. 
 
 The majority of the trends that can be observed follow what one can expect from common sense. Here, we will just highlight a few of the conclusions that can be draw from studying these figures:
\begin{enumerate}
\item Decreasing the size of the units improves the direction reconstruction, mainly at the high end of the energy spectrum. However, the advantage of going from 4~m$\times$4~m to  2~m$\times$2~m (or smaller) is marginal. Therefore, designing detector units smaller than  4~m$\times$4~m is not the most economical way to improve on angular resolution. 
\item An array sizes below 100~m$\times$100~m need to be avoided since $\gamma$-ray efficiency, angular resolution, and proton rejection power become all very poor. The most likely reason for this is that size of air showers at all energies are comparable to the size of the array and are therefore only partially contained in the array leading to poor performance.
\item Decreasing {\it{unit threshold}} and {\it trigger multiplicity} increases the $\gamma$-ray efficiency at low energies, however this comes with a decrease of angular resolution at the same energy range. Optimization should therefore be performed carefully, taking into account the angular resolution needed for a specific scientific goals (like Gamma-Ray-Burst detection).  However, designing a smart trigger algorithm, allowing triggering into the $\sim$100\,GeV energy range, might be a cost efficient way of extending the lower energy reach of the observatory. Lowering the {\it {unit threshold}} below 10 MeV gives only marginal improvements and should therefore not be pursued in detector design. 
\item The largest, fully filled arrays at the highest elevation gives the best performance in $\gamma$-ray detection efficiencies. However, elevation does only slightly improve the proton rejection efficiency using muons while increasing the array-size up to 400~m$\times$400~m still increases the proton rejections power. When deciding on a site for an observatory, one should consider the option of building a larger detector for the same price at lower altitude if this allows for an improved performance.
\end{enumerate}

\section{Discussion \& conclusion}
We presented here the trade-off on performance for design choices for ASPD observatories. We tried to select a few key design parameters and studied figures of merit for the performance of such an observatory. Although, the parameter space is large and multi-dimensional, we were able to identify a few key guidelines when designing an ASPD observatory without the need for dedicated hardware simulations. 

The analysis presented here does not fully address all the issues that one needs to consider before designing an $\gamma$-ray ASDP observatory. Trigger multiplicities where chosen pragmatically at fixed values, however, in reality this parameter is driven by trigger rate, data throughput, data reduction and data storage capabilities of the observatory. In order to address this properly, a realistic noise model (including single muons and small air showers) needs to be developed. This will be part of future work. In addition, for the reduction of hadronic induced EAS we used a pragmatic approach of having an ideal muon detection system. Such a system shows very good proton rejection capabilities and should therefore be considered in a future observatory. However, there are other parameters \cite{HAWC_CRAB} that will also help by the identification of hadronic induced showers. In future work we will incorporate a few more parameters and study how they perform individually and in concert.

\section*{Acknowledgements}
The Authors would like to express their gratitude to the HAWC collaboration for many fruitful discussions and feedback.

\bibliographystyle{JHEP}
\bibliography{references}
%\bibliography{references}

%\begin{thebibliography}{99}
%\bibitem[HESS]{Some HESS citations}
%\bibitem[VERITS]{Some HESS citations}
%\bibitem[CTA]{Some HESS citations}
%\bibitem[HAWC]{Some HESS citations}....
%
%\end{thebibliography}

\end{document}